\documentclass[twocolumn,showpacs,preprintnumbers,superscriptaddress, amsmath,amssymb]{revtex4}
\usepackage{amsthm}
\usepackage{bm}
\usepackage{epsfig}

\DeclareFontFamily{OT1}{rsfs}{}
\DeclareFontShape{OT1}{rsfs}{m}{n}{<5> rsfs5 <7> rsfs7 <10> rsfs10}{}
\DeclareSymbolFont{mathrsfs}{OT1}{rsfs}{m}{n}
\DeclareSymbolFontAlphabet{\mathrsfs}{mathrsfs}

\renewcommand*{\Im}{{\rm Im}\,}
\renewcommand*{\Re}{{\rm Re}\,}

\newcommand{\A}{\mathbf{A}}
\newcommand{\K}{\mathbf{k}}
\renewcommand{\P}{\mathbf{p}}
\newcommand{\R}{\mathbf{r}}

\newcommand{\kk}{{k_{\parallel}}}
\renewcommand{\k}{{k_{\perp}}}
\newcommand{\pp}{{p_{\parallel}}}
\newcommand{\p}{{p_{\perp}}}

\newcommand*{\arccosh}{{\rm arccosh}\,}

\renewcommand{\H}{\hat{H}}

\begin{document}

\title{Non-sequential double ionization below laser-intensity threshold: Anticorrelation of electrons without excitation of parent ion}

\author{D. I. Bondar}
\email{dbondar@sciborg.uwaterloo.ca}
\affiliation{University of Waterloo, Waterloo, Ontario N2L 3G1, Canada}
\affiliation{National Research Council of Canada, Ottawa, Ontario K1A 0R6, Canada}

\author{G. L. Yudin}
\email{gennady.yudin@nrc.ca}
\affiliation{National Research Council of Canada, Ottawa, Ontario K1A 0R6, Canada}
\affiliation{Universit\'{e} de Sherbrooke, Sherbrooke, Qu\'{e}bec J1K 2R1, Canada}

\author{W.-K. Liu}
\affiliation{University of Waterloo, Waterloo, Ontario N2L 3G1, Canada}

\author{M. Yu. Ivanov}
\affiliation{Imperial College, London SW7 2BW, UK}

\author{A. D. Bandrauk}
\affiliation{Universit\'{e} de Sherbrooke, Sherbrooke, Qu\'{e}bec J1K 2R1, Canada}

\begin{abstract}
Two-electron correlated spectra of non-sequential double ionization below laser-intensity threshold are known to exhibit back-to-back scattering of the electrons, viz., the anticorrelation of the electrons. Currently, the widely accepted interpretation of the anticorrelation is recollision-induced excitation of the ion plus subsequent field ionization of the second electron. We argue that another mechanism, namely simultaneous electron emission, when the time of return of the rescattered electron is equal to the time of liberation of the bounded electron (the ion has no time for excitation), can also explain the anticorrelation of the electrons in the deep below laser-intensity threshold regime. Our conclusion is based on the results of the numerical solution of the time-dependent Schr\"{o}dinger equation for a model system of two one-dimensional electrons as well as an adiabatic analytic model that allows for a closed-form solution.
\end{abstract}


\maketitle

\section{Introduction} 

After one-electron ionization of an atom or a molecule, the free electron can recollide with its parent ion \cite{Kuchiev_1987, Corkum_1993}. Inelastic scattering on the ion may result in collisional excitation or ionization. Liberation of the second electron during recollision is known as non-sequential double ionization (NSDI). 

Originally, NSDI has been discovered experimentally for alkaline-earth atoms \cite{Suran1975} (regarding further experimental investigations of NSDI for alkaline- earth atoms, see, e.g., Refs. \cite{Bondar1993, Bondar1998, Bondar2000, Liontos2004, Liontos2008, Liontos2010} and references therein). Later, NSDI has been found in nobel gas atoms \cite{LHuillier1982, LHuillier1983}.

According to classical calculations, the maximum kinetic energy of the recolliding electron is $\approx 3.2 U_p$ \cite{Corkum_1993}, where
$U_p=(F/2\omega)^2$, $F$ is the laser field strength, and $\omega$
is the laser frequency [unless stated otherwise, the Hartree atomic units (a.u., $\hbar=m_e=|e|=1$) are used throughout the paper]. Hence, one can divide NSDI into two types: when the energy of the recolliding electron is enough to collisionally ionize the parent ion -- the above laser-intensity threshold regime, and when the kinetic energy is insufficient to directly ionize the ion -- the below laser-intensity threshold (BIT) regime. The former one regime is thoroughly studied experimentally as well as theoretically (see, e.g., Refs. \cite{Lein2000, Yudin_2001A, Panfili_2002, deMorissonFaria2003, deMorissonFaria2004, Phay2005,  Becker2005a, deMorissonFaria2005, Ho_2005, Ho2006, Haan2006, deMorissonFaria2006, Liu_2006, Taylor2007, Staudte_2007, Rudenko_2007, deMorissonFaria2008b, Zhou2010, Chen2010} and references therein).

Despite tremendous challenges, NSDI BIT has been experimentally observed \cite{Weckenbrock_2004, Rudenko_2004, Zeidler_2005, Zrost_2006, Liu2008a, Liu2010}. The main feature of this regime, revealed by measured correlated two-electron spectra, is that the two electrons prefer to drift out in opposite directions, manifesting the so-called anticorrelation of the electrons. The interpretation of this observation is the aim of recent theoretical studies \cite{Haan2008a, Emmanouilidou2009, Bondar2009b, Shaaran2010, Shaaran2010a, Ye2010, Haan2010}. Briefly summarizing these models,  it is widely accepted that recollision-induced excitation of the ion plus subsequent field ionization of the second electron (RESI) is a main mechanism of NSDI BIT. Nonetheless, alternative, non-RESI, mechanisms are also considered (see, e.g., Refs. \cite{Parker2006, Liu2008a, Ye2010, Chen2010} and references therein).

However, is excitation of the parent ion indeed necessary to explain the electron anticorrelation in NSDI BIT? We show that this is not always the case. Our conclusions are based, first, on model {\it ab-initio} calculations for a system of two one-dimensional electrons show that the anticorrelation of the electrons exists even if the ion has only a single bound state (Sec. \ref{Sec2}). Second, within the adiabatic approximation (see, e.g., Refs. \cite{Dykhne_1962, Solovev1976, Solovev1989, Savichev1991a, Solovev2005, Tolstikhin2008a, Bondar2009, Tolstikhin2010A, Batishchev2010} and references therein), we present a simple analytical three-dimensional model based on the assumption that both the electrons are ejected simultaneously (the time of return of the first electron coincides with the time of liberation of the second electron, i.e., the ion has no time to be excited) in Sec. \ref{Sec3}. An advantage of this model is that it allows for a simple analytical solution in closed form. In a certain range of parameters, the correlated two-electron spectrum obtained within this model exhibits the anticorrelation of the electrons. This mechanism of simultaneous electron emission can produce the anticorrelation of the electrons in the deep BIT regime.

\section{Ab inito evidence of the new mechanism}\label{Sec2}

Consider the model Hamiltonian for a system of two one-dimensional electrons 
\begin{eqnarray}\label{Model_Hamiltonian}
\H(t) &=& \left( \hat{p}_1^2 + \hat{p}_2^2 \right)/2 + V(x_1)  + V(x_2) \nonumber\\
&& - V(x_2 - x_1) + (x_1 + x_2) F_L(t),
\end{eqnarray}
where $V(x) =  -4\exp(-3x^2)$ is the prototype for the potential of the electron-core attraction, $-V(x_2 - x_1)$ is the prototype of the electron-electron repulsion, and $F_L(t) = F f(t) \sin (\omega t)$ ($F=0.05$ and $\omega = 0.6$) -- the laser pulse, where $f(t)$ is a trapezoid with one-cycle turn-on, six-cycle full strength, and one-cycle turn-off. The potential $V(x)$ is chosen such that the one-particle Hamiltonian, $\hat{H}_{ion} = -\partial^2/(2 \partial x^2) + V(x)$, supports only one bound state with the ionization potential $I_p = 2.11$. 

To make analysis more transparent, we substitute the original problem (NSDI) by the corresponding problem of laser-assisted scattering, i.e., we simply discard the first step of NSDI -- liberation of the first electron. In other words, instead of assuming that the two-electron system initially is in the ground state, we assume that the first electron is an incident wave-packet and the second electron is in the single bound state of the ion. This modification is very useful, as it allows us to eliminate all other possible interactions and processes except the three major components -- the electron-electron repulsion, the electron-core attraction, and the electron-laser field interaction. 

We solve numerically, by means of the split-operator method, the time-dependent Schr\"{o}dinger equation,
\begin{eqnarray}\label{SchrodingerEq_ModelWaveFunc}
i\partial \Psi (x_1, x_2; t) /\partial t = \H(t)\Psi (x_1, x_2; t). 
\end{eqnarray}
The initial unsymmetrized wave function takes the form
\begin{eqnarray}
\Psi (x_1, x_2; 0) = \exp\left[  -\frac{(x_1- \mu)^2}{(2\sigma_x)^2}  -i\sqrt{2E_{in}}x_1 \right]\psi ( x_2 ),
\end{eqnarray}
where $\sigma_x = 2$, $\mu = 7\sigma_x$, $\hat{H}_{ion} \psi ( x ) = -I_p \psi ( x )$, and $E_{in} = 0, 0.3$ is the mean value of the kinetic energy of the incident electron. The parameters are selected such that,
$
 I_p - (F/\omega)^2/2 \approx 4 \omega
$
and 
$
I_p - (\sqrt{0.6} + F/\omega)^2 /2 \approx 3 \omega
$, i.e., the bound electron needs to absorb at least four (three) photons to be liberated in the case of $E_{in}=0$ ($E_{in}=0.3$). The wave function then is symmetrized, i.e., we assume that the spins of the electrons are antiparallel.

\begin{figure}
\begin{center}
\includegraphics[scale=0.5]{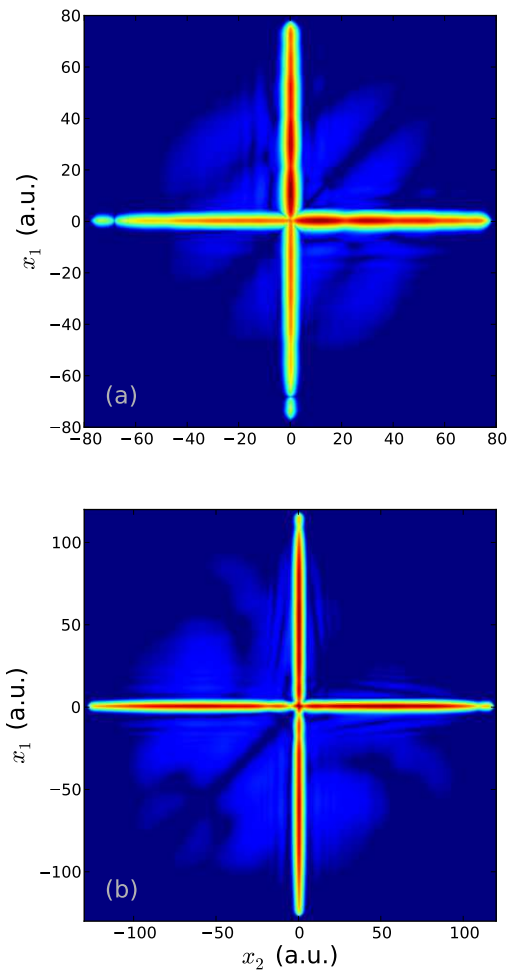}
\end{center}
\caption{(Color online) Plots of the wave functions of the model system, $\log \left |\Psi (x_1, x_2; 73) \right|^2$, for different values of the kinetic energy of the incident electron [Eqs. (\ref{Model_Hamiltonian}) and (\ref{SchrodingerEq_ModelWaveFunc})]. (a) $E_{in}=0$; (b) $E_{in}=0.3$.}\label{wavefunctions_Fig}
\end{figure}

The wave function of the model system is pictured in Fig. \ref{wavefunctions_Fig} for two values of $E_{in}$.  The wave function in the absence of the laser produces only maxima on the axes $x_{1,2}$ corresponding to one electron bound and the other one free; hence, this part of the wave function, which also shows up when the laser field is turned on, should be ignored because it does not correspond to NSDI. When the mean kinetic energy of the incident particle is zero, Fig. \ref{wavefunctions_Fig}(a), we observe that the two electrons ``prefer'' to be anticorrelated rather than correlated. However, once the mean kinetic energy is slightly increased such that NSDI still progresses in the BIT regime [see Fig. \ref{wavefunctions_Fig}(b)], the electron correlation begins to dominate the anticorrelation. Note that the wave function in the third quadrant of Fig. \ref{wavefunctions_Fig}(b) (i.e., the portion of the wave function that describes the correlation of the electrons) has a characteristic V-like (aka fingerlike) shape, which is a dominant shape of the two-electron correlated spectra of NSDI in the above threshold regime.

\section{The simultaneous electron emission within the adiabatic approximation}\label{Sec3}

We calculate the correlated spectrum within the adapted Savichev  adiabatic approximation \cite{Savichev1991a, Bondar2009}
\begin{eqnarray}\label{AdiabaticIonizRates}
\Gamma \propto \exp\left( -\frac2{\omega} \Im\int^{\varphi_0} [E_f (\varphi)- E_i(\varphi)] d\varphi \right),
\end{eqnarray}
which is a generalization of the Dykhne adiabatic method \cite{Dykhne_1962} to the case when a quantum system has a continuum spectrum. Here, $\varphi = \omega t$ is the phase of a laser field, $\varphi_0$ is the complex solution of the equation $E_i(\varphi_0) = E_f(\varphi_0)$ with the smallest positive imaginary part, $E_{i,f}(\varphi)$ being the adiabatic terms of the system ``before'' and ``after'' a non-adiabatic transition. If the quasi classical approximation to $\Gamma$ is sufficient, then instead of using the adiabatic terms, one can employ the total energies of the corresponding classical systems \cite{Bondar2009}.

As previously, we replace the problem of NSDI BIT by the problem of laser-assisted scattering of an electron by an ion. We study the simultaneous electron emission (SEE) process, when the moment of collision of the incident electron coincides with the moment of ionization of the ion.

To include electron-electron repulsion, we apply the standard exponential perturbation theory. Namely, first we find electron  energies and trajectories without the electron-electron repulsion. Then, we correct electron action and energies by adding the effect of the electron-electron repulsion, which calculated along the zero-order trajectories.

Thus, in zero order, we define $E_{i,f}(\varphi)$ without the electron-electron repulsion. Since before ionization, one electron is free and the other is bound, the total classical energy of the system before collision is
\begin{eqnarray}\label{InitialEnergy}
E_i(\varphi) &=& \left[ \P + \A(\varphi) \right]^2/2 - I_p, 
\end{eqnarray}
where $\P$ is the (three-dimensional) canonical momentum (i.e., the kinetic momenta at $\varphi = \pm\infty$) of the incident electron, $I_p$ is the ionization potential of the ion, and 
$
\A(\varphi) = -({\bf F}/\omega) \sin\varphi
$
is the vector potential of a linearly polarized laser field. After collision both the electrons are free, and the classical energy of the system reads
\begin{eqnarray}\label{FinalEnergy}
E_f(\varphi) &=&  \left[ \K_1 + \A(\varphi) \right]^2/2 + \left[ \K_2 + \A(\varphi) \right]^2/2,
\end{eqnarray}
where $\K_{1,2}$ are canonical momenta of the first and second electrons.

Such two electron process is formally equivalent to single-electron strong field ionization of a quasiatom (within a pre-exponential accuracy). This statement is manifested by the following equality 
\begin{eqnarray}\label{SingleElectronIonizationAndSEE}
&& \frac 1{\omega} \int^{\varphi_0} \left[ E_f(\varphi) - E_i(\varphi) \right]d\varphi \nonumber\\
&& \quad = \frac 1{\omega} \int^{\varphi_0} \left\{ \frac 12 \left[ \mathrsfs{K} + A(\varphi) \right]^2 + \mathrsfs{I}_p \right\}d\varphi,
\end{eqnarray}
where the right hand side of Eq. (\ref{SingleElectronIonizationAndSEE}) is the action of single-electron ionization within the strong field approximation, $\mathrsfs{K} = \kk_1 + \kk_2 - \pp$ is the effective longitudinal momentum, and  $\mathrsfs{I}_p = I_p + \left(  \K_1^2 + \K_2^2 - \P^2 - \mathrsfs{K}^2 \right)/2$ 
is the effective ionization potential of the quasiatom. (Note that such a reduction is in fact more general and can be applied to an arbitrary number of electrons, for further discussions see Ch. 7 of Ref. \cite{Bondar2010}.) Therefore, derivation of the correlated spectra for SEE is reduced to calculation of the momentum distribution of photoelectrons after single-electron ionization without any restrictions on $\mathrsfs{K}$. The most suitable solution of the last problem for our current discussion has been found in Ref. \cite{Bondar2008}. Note that the equation obtained in Ref. \cite{Bondar2008} is valid only for positive $\mathrsfs{I}_p$; hence, the case of $\mathrsfs{I}_p < 0$ being of interest for SEE, must also be considered.

\begin{widetext}
Substituting Eqs. (\ref{InitialEnergy}) and (\ref{FinalEnergy}) into Eq. (\ref{AdiabaticIonizRates}) and taking into account the previous comments, we obtain
\begin{eqnarray}
& \Gamma (\K_1, \K_2) \propto \exp\left( -2 \left|\mathrsfs{I}_p\right| G /\omega + V_{ee} \right), \label{QuasiClassicalCorrSpectra} &\\
& \varphi_0 = \left\{
\begin{array}{ccc}
\arcsin x_- + i\arccosh x_+ &\mbox{if}& \mathrsfs{I}_p < 0 \\
 \arcsin y_- + i\arccosh y_+ &\mbox{if}&  \mathrsfs{I}_p > 0,
\end{array}
\right. & \\
& G = \left\{
\begin{array}{r}
\left( \eta^2 + \frac 1{2g^2} - 1 \right) \arccosh x_+ - \sqrt{x_+^2 - 1} 
	 \left( \frac{2\eta x_-}g + x_+ \frac{1-2x_-^2}{2g^2} \right)  \mbox{ if } \mathrsfs{I}_p < 0, \\
\left( \eta^2 + \frac 1{2g^2}  +1 \right) \arccosh y_+ - \sqrt{y_+^2 - 1} 
	\left( \frac{2\eta y_-}g + y_+ \frac{1-2y_-^2}{2g^2}\right)  \mbox{ if }  \mathrsfs{I}_p > 0, 
\end{array}
\right. & \\
& x_{\pm} = \left| g(\eta-1) + 1\right|/2 \pm \left| g(\eta-1) - 1 \right|/2, & \nonumber\\
& y_{\pm} = \sqrt{ (g\eta+1)^2 + g^2}/2 \pm \sqrt{ (g\eta -1)^2 + g^2 }/2 , & \nonumber\\
& g = \omega \sqrt{2\left|\mathrsfs{I}_p\right|} / F, \qquad \eta = \mathrsfs{K} / \sqrt{2\left|\mathrsfs{I}_p\right|}.\nonumber &
\end{eqnarray}
In Eq. (\ref{QuasiClassicalCorrSpectra}), $V_{ee}$ denotes a crucial correction due to the electron-electron repulsion, 
\begin{eqnarray}
V_{ee} = -\frac 2{\omega} \Im \int_{\Re\varphi_0}^{\varphi_0} \frac{d\varphi}{\left| \R_1(\varphi) - \R_2 (\varphi) \right|} 
= \lim_{\varepsilon\to 0} \frac{-2}{\left| \K_1 - \K_2 \right|} \Im\int_{\Re\varphi_0}^{\varphi_0} \frac{d\varphi}{\sqrt{(\varphi-\varphi_0)^2 + \varepsilon^2}}
= -\pi / \left| \K_1 - \K_2 \right| .
\end{eqnarray}
\end{widetext}

The contribution of the Coulomb potential has been calculated along the field-free trajectories \cite{Smirnova_2008, Bondar2009b}
$$
\R_{1,2}(\varphi) = \frac 1{\omega} \int_{\varphi_0}^{\varphi} \left[\K_{1,2} + \A(\phi)\right] d\phi.
$$
The condition of validity of Eq. (\ref{QuasiClassicalCorrSpectra}) is $\Gamma \ll 1$, which is merely the condition of applicability of the adiabatic approximation.

\begin{figure}
\begin{center}
\includegraphics[scale=0.5]{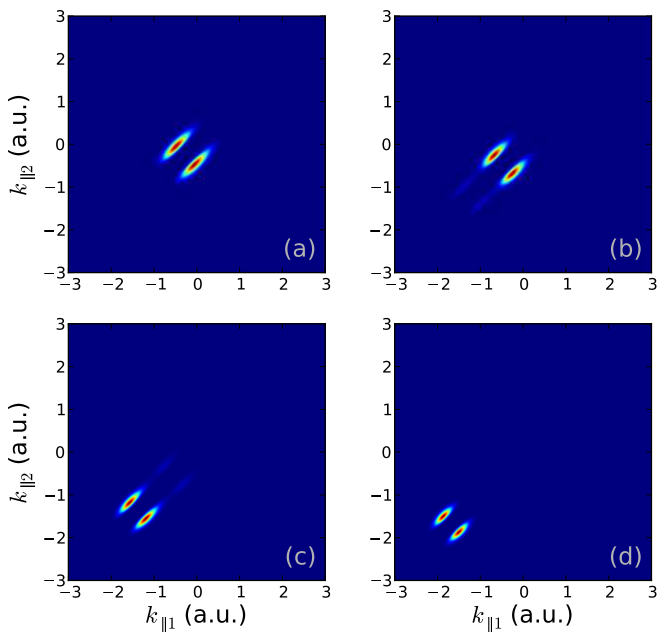}
\end{center}
\caption{(Color online) Correlated two-electron spectra of Ar (linear scale) given by Eq. (\ref{QuasiClassicalCorrSpectra}) at  800 nm and $1\times 10^{13}$ W/cm$^2$ ($\p = \k_1 = \k_2 = 0$) for different momenta of the incident electron: (a) $\pp = 0.1$, (b) $\pp = 0.2$, (c) $\pp=0.25$, (d) $\pp=0.4$.}\label{IncidentMomentumDependence}
\end{figure}

\begin{figure}
\begin{center}
\includegraphics[scale=0.5]{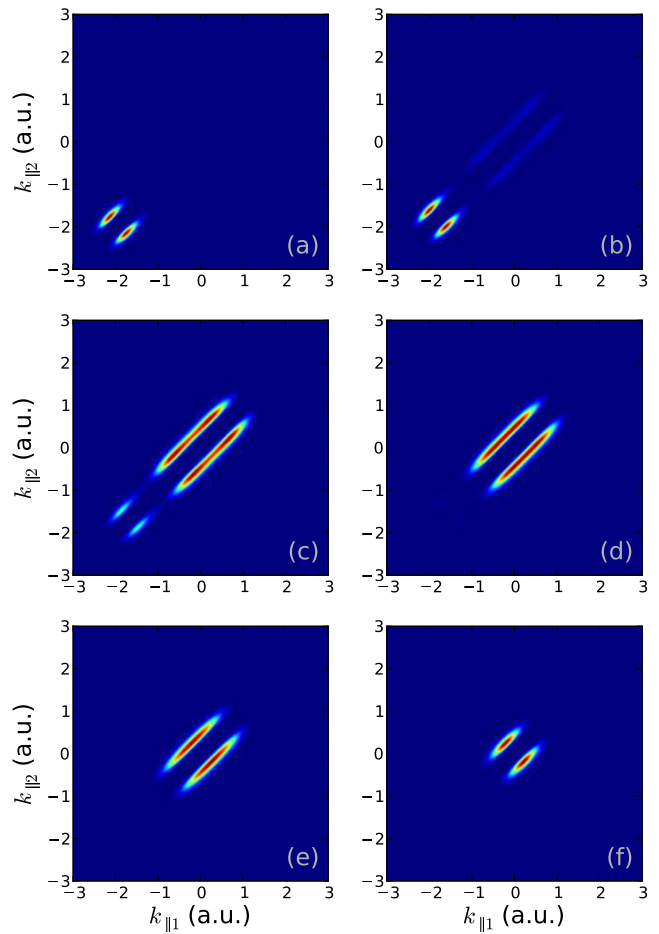}
\end{center}
\caption{(Color online) Correlated two-electron spectra of Ar (linear scale) given by Eq. (\ref{QuasiClassicalCorrSpectra}) at  800 nm ($\P = {\bf 0}$ and $\k_1 = \k_2 = 0$) for different intensities of the laser field: (a) $3\times 10^{13}$ W/cm$^2$, (b) $2.7\times 10^{13}$ W/cm$^2$, (c) $2.5\times 10^{13}$ W/cm$^2$, (d) $2.3\times 10^{13}$ W/cm$^2$, (e) $2\times 10^{13}$ W/cm$^2$, and (f) $1\times 10^{13}$ W/cm$^2$.}\label{IntensityDependence}
\end{figure}

The model of NSDI BIT presented here [Eq. (\ref{QuasiClassicalCorrSpectra})] is similar to the one reported in Ref. \cite{Bondar2009b}. However, there is an important difference. In the correlated spectra (\ref{QuasiClassicalCorrSpectra}), the momentum of the incident electron ${\bf p}$ is a free parameter, whereas the correlated spectrum given by Eq. (19) of Ref. \cite{Bondar2009b} does not have this freedom -- the canonical momentum of the recolliding electron is fixed as a function of the phase of recollision (this functional dependence is obtained from the saddle-point analysis of the S-matrix amplitude for NSDI). Therefore, Eq. (\ref{QuasiClassicalCorrSpectra}) allows us to establish links between values of the canonical momentum of the rescattered (incident) electron and specific portions of the correlated spectra. The dependance of the correlated spectra (\ref{IncidentMomentumDependence}) on the momentum of the incident electron is presented in Fig. \ref{IncidentMomentumDependence}.

The peaks of the momentum distributions depicted in Figs. \ref{IncidentMomentumDependence}(c) and \ref{IncidentMomentumDependence}(d) are located approximately at $10\sqrt{U_p}$. This value  is much larger than $2\sqrt{U_p}$ that is the classical drift momentum of a free electron in a laser field. Such a discrepancy is due to the fact that the momentum of the incident electron ${\bf p}$ can take an arbitrary value in our model, and its value is not determined by the saddle-point integration of the corresponding S-matrix amplitude, which is in turn equivalent to the employment of the quasi-classical description of the rescattered electron. (As we have mentioned, the incident electron momentum was fixed in Ref. \cite{Bondar2009b}, and hence, such high electron momenta were not observed.)

Comparing Figs.  \ref{wavefunctions_Fig} and \ref{IncidentMomentumDependence}, one concludes that the analytical model [Eq. (\ref{QuasiClassicalCorrSpectra})] qualitatively agrees with the numerical solution of the Schr\"{o}dinger equation (\ref{SchrodingerEq_ModelWaveFunc}) for the model system of two one-dimensional electrons presented in Sec. \ref{Sec2}. Indeed, when the momentum of the incident electron is zero -- the anticorrelation of the electrons dominates the correlation [Figs. \ref{IncidentMomentumDependence}(a)]. The situation is opposite when the momentum of the incident electron is slightly increased [Figs. \ref{IncidentMomentumDependence}(c) and \ref{IncidentMomentumDependence}(d)].

An interesting question is the dependence of the correlated spectrum (\ref{QuasiClassicalCorrSpectra}) on the intensity of the laser field, which is pictured in Fig. \ref{IntensityDependence}. The parameters used to plot Fig. \ref{IntensityDependence}(a) coincide with the parameters employed in the recent experiment \cite{Liu2010}. Yet, the experimentally measured correlated spectrum exhibits peaks in the second and fourth quadrants, which contradicts Fig. \ref{IntensityDependence}(a). The reason of such a disagreement is that the anticorrelation in these experimental data is due to the RESI mechanism. As the intensity lowers, the peaks in the correlated spectra of SEE shift to the second and fourth quadrants [see Fig. \ref{IntensityDependence}(f)], i.e., the SEE process leeds to the anticorrelation of the electrons in the deep BIT regime. This can be explain intuitively in the following way: The lower the intensity, the smaller the canonical momentum of the returning (incident) electron. Since the canonical momentum of the system is approximately conserved, we obtain ${\bf 0} \approx \K_1 + \K_2$; hence, $\K_1 \approx -\K_2$. Having absorbed a necessary number of photons, both the electrons emerge in the continuum where they experience strong electron-electron repulsion that pushes them apart. Therefore, during SEE, the electrons gain momenta because of the electron-electron repulsion. Indeed, if the electron-electron interaction is ``turned off''  in Eq. (\ref{QuasiClassicalCorrSpectra}) by setting $V_{ee} = 0$, then we obtain a single peak centred at the origin instead of the two peaks visible in Fig. \ref{IntensityDependence}(f). Recall that contrary to SEE, the electrons gain kinetic energy by oscillating in the laser field in the course of RESI.

\section{Conclusions and discussions} 

We have demonstrated that the mechanism of simultaneous electron emission, when the time of return of the rescattered electron is equal to the time of liberation of the bound electron, can be responsible for the anticorrelation of the electrons during NSDI in the deep BIT regime [see Fig. \ref{IntensityDependence}(f)]. The SEE process significantly differs from RESI because it does not require the excitation of the ion to explain the anticorrelation of the electrons observed in the two-electron correlated spectra. According to SEE, NSDI BIT can be realized by a rescattered electron with a very low (canonical) momentum. Such an electron located near the parent ion creates strong Coulomb field that acts on the bound electron. This Coulomb repulsion eventually ionizes the parent ion and leads to the anticorrelation of both the electrons. The SEE should be manifested at small laser intensities because the lower the strength of the laser field, the lower the canonical momentum of the rescattered electron.

Another important difference between SEE and RESI mechanisms is that the former results strictly in the anticorrelation of the electrons, viz., the correlated two-electron spectra should be concentrated mainly in the second and fourth quadrants. Whereas, the RESI mechanism leads to electron momentum distributions more or less equally distributed in all four quadrants of the momentum plane \cite{Shaaran2010, Shaaran2010a}.

The main difference between SEE and other proposed non-RESI mechanisms (see, e.g., Refs. \cite{Parker2006, Liu2008a, Ye2010}) is that it is induced by rescattered electrons with nearly zero momentum, and hence it requires no time delay between the return of the first electron and liberation of the second. 

SEE and RESI are by no means mutually exclusive processes of NSDI; they both contribute. It is important to study quantitatively the relative contribution of SEE and RESI to NSDI in the deep BIT regime.  Nevertheless, this cannot be done by neither the reduced dimensional numerical model presented in Sec. \ref{Sec2}, nor by the analytical adiabatic model developed in Sec. \ref{Sec3}.

Our analytical model can be further developed in the spirit of the correlated Keldysh-Faisal-Reiss theory \cite{Becker1994, Becker1995, Golovinski1997}.
However, if one wants to study NSDI in very weak laser fields, the model may need a substantial revision because in such a regime, the Coulomb interaction dominants over  the interaction with the laser, as a result, the classical trajectories of the electrons after ionization become chaotic (see, e.g., Refs. \cite{Chirikov1979, Krainov2010}).

\acknowledgments 

D.I.B. thanks the Ontario Graduate Scholarship program for financial support.  M.Yu.I. and W.-K.L. acknowledge support of NSERC discovery grants.

\bibliography{Novel_NSDI}

\begin{thebibliography}{61}
\expandafter\ifx\csname natexlab\endcsname\relax\def\natexlab#1{#1}\fi
\expandafter\ifx\csname bibnamefont\endcsname\relax
  \def\bibnamefont#1{#1}\fi
\expandafter\ifx\csname bibfnamefont\endcsname\relax
  \def\bibfnamefont#1{#1}\fi
\expandafter\ifx\csname citenamefont\endcsname\relax
  \def\citenamefont#1{#1}\fi
\expandafter\ifx\csname url\endcsname\relax
  \def\url#1{\texttt{#1}}\fi
\expandafter\ifx\csname urlprefix\endcsname\relax\def\urlprefix{URL }\fi
\providecommand{\bibinfo}[2]{#2}
\providecommand{\eprint}[2][]{\url{#2}}

\bibitem[{\citenamefont{Kuchiev}(1987)}]{Kuchiev_1987}
\bibinfo{author}{\bibfnamefont{M.}~\bibnamefont{Kuchiev}},
  \bibinfo{journal}{JETP Letters} \textbf{\bibinfo{volume}{45}},
  \bibinfo{pages}{404} (\bibinfo{year}{1987}).

\bibitem[{\citenamefont{Corkum}(1993)}]{Corkum_1993}
\bibinfo{author}{\bibfnamefont{P.~B.} \bibnamefont{Corkum}},
  \bibinfo{journal}{Phys. Rev. Lett.} \textbf{\bibinfo{volume}{71}},
  \bibinfo{pages}{1994} (\bibinfo{year}{1993}).

\bibitem[{\citenamefont{Suran and Zapesochnyi}(1975)}]{Suran1975}
\bibinfo{author}{\bibfnamefont{V.~V.} \bibnamefont{Suran}} \bibnamefont{and}
  \bibinfo{author}{\bibfnamefont{I.~P.} \bibnamefont{Zapesochnyi}},
  \bibinfo{journal}{Sov. Tech. Phys. Lett.} \textbf{\bibinfo{volume}{1}},
  \bibinfo{pages}{420} (\bibinfo{year}{1975}).

\bibitem[{\citenamefont{Bondar' and Suran}(1993)}]{Bondar1993}
\bibinfo{author}{\bibfnamefont{I.~I.} \bibnamefont{Bondar'}} \bibnamefont{and}
  \bibinfo{author}{\bibfnamefont{V.~V.} \bibnamefont{Suran}},
  \bibinfo{journal}{JETP} \textbf{\bibinfo{volume}{76}}, \bibinfo{pages}{381}
  (\bibinfo{year}{1993}).

\bibitem[{\citenamefont{Bondar' and Suran}(1998)}]{Bondar1998}
\bibinfo{author}{\bibfnamefont{I.~I.} \bibnamefont{Bondar'}} \bibnamefont{and}
  \bibinfo{author}{\bibfnamefont{V.~V.} \bibnamefont{Suran}},
  \bibinfo{journal}{JETP Letters} \textbf{\bibinfo{volume}{68}},
  \bibinfo{pages}{837} (\bibinfo{year}{1998}).

\bibitem[{\citenamefont{Bondar' et~al.}(2000)\citenamefont{Bondar', Suran, and
  Dudich}}]{Bondar2000}
\bibinfo{author}{\bibfnamefont{I.~I.} \bibnamefont{Bondar'}},
  \bibinfo{author}{\bibfnamefont{V.~V.} \bibnamefont{Suran}}, \bibnamefont{and}
  \bibinfo{author}{\bibfnamefont{M.~I.} \bibnamefont{Dudich}},
  \bibinfo{journal}{J. Phys. B} \textbf{\bibinfo{volume}{33}},
  \bibinfo{pages}{4243} (\bibinfo{year}{2000}).

\bibitem[{\citenamefont{Liontos et~al.}(2004)\citenamefont{Liontos, Bolovinos,
  Cohen, and Lyras}}]{Liontos2004}
\bibinfo{author}{\bibfnamefont{I.}~\bibnamefont{Liontos}},
  \bibinfo{author}{\bibfnamefont{A.}~\bibnamefont{Bolovinos}},
  \bibinfo{author}{\bibfnamefont{S.}~\bibnamefont{Cohen}}, \bibnamefont{and}
  \bibinfo{author}{\bibfnamefont{A.}~\bibnamefont{Lyras}},
  \bibinfo{journal}{Phys. Rev. A} \textbf{\bibinfo{volume}{70}},
  \bibinfo{pages}{033403} (\bibinfo{year}{2004}).

\bibitem[{\citenamefont{Liontos et~al.}(2008)\citenamefont{Liontos, Cohen, and
  Bolovinos}}]{Liontos2008}
\bibinfo{author}{\bibfnamefont{I.}~\bibnamefont{Liontos}},
  \bibinfo{author}{\bibfnamefont{S.}~\bibnamefont{Cohen}}, \bibnamefont{and}
  \bibinfo{author}{\bibfnamefont{A.}~\bibnamefont{Bolovinos}},
  \bibinfo{journal}{J. Phys. B} \textbf{\bibinfo{volume}{41}},
  \bibinfo{pages}{045601} (\bibinfo{year}{2008}).

\bibitem[{\citenamefont{Liontos et~al.}(2010)\citenamefont{Liontos, Cohen, and
  Lyras}}]{Liontos2010}
\bibinfo{author}{\bibfnamefont{I.}~\bibnamefont{Liontos}},
  \bibinfo{author}{\bibfnamefont{S.}~\bibnamefont{Cohen}}, \bibnamefont{and}
  \bibinfo{author}{\bibfnamefont{A.}~\bibnamefont{Lyras}}, \bibinfo{journal}{J.
  Phys. B} \textbf{\bibinfo{volume}{43}}, \bibinfo{pages}{095602}
  (\bibinfo{year}{2010}).

\bibitem[{\citenamefont{L'Huillier et~al.}(1982)\citenamefont{L'Huillier,
  Lompre, Mainfray, and Manus}}]{LHuillier1982}
\bibinfo{author}{\bibfnamefont{A.}~\bibnamefont{L'Huillier}},
  \bibinfo{author}{\bibfnamefont{L.~A.} \bibnamefont{Lompre}},
  \bibinfo{author}{\bibfnamefont{G.}~\bibnamefont{Mainfray}}, \bibnamefont{and}
  \bibinfo{author}{\bibfnamefont{C.}~\bibnamefont{Manus}},
  \bibinfo{journal}{Phys. Rev. Lett.} \textbf{\bibinfo{volume}{48}},
  \bibinfo{pages}{1814} (\bibinfo{year}{1982}).

\bibitem[{\citenamefont{L'Huillier et~al.}(1983)\citenamefont{L'Huillier,
  Lompre, Mainfray, and Manus}}]{LHuillier1983}
\bibinfo{author}{\bibfnamefont{A.}~\bibnamefont{L'Huillier}},
  \bibinfo{author}{\bibfnamefont{L.~A.} \bibnamefont{Lompre}},
  \bibinfo{author}{\bibfnamefont{G.}~\bibnamefont{Mainfray}}, \bibnamefont{and}
  \bibinfo{author}{\bibfnamefont{C.}~\bibnamefont{Manus}},
  \bibinfo{journal}{Phys. Rev. A} \textbf{\bibinfo{volume}{27}},
  \bibinfo{pages}{2503} (\bibinfo{year}{1983}).

\bibitem[{\citenamefont{Lein et~al.}(2000)\citenamefont{Lein, Gross, and
  Engel}}]{Lein2000}
\bibinfo{author}{\bibfnamefont{M.}~\bibnamefont{Lein}},
  \bibinfo{author}{\bibfnamefont{E.~K.~U.} \bibnamefont{Gross}},
  \bibnamefont{and} \bibinfo{author}{\bibfnamefont{V.}~\bibnamefont{Engel}},
  \bibinfo{journal}{Phys. Rev. Lett.} \textbf{\bibinfo{volume}{85}},
  \bibinfo{pages}{4707} (\bibinfo{year}{2000}).

\bibitem[{\citenamefont{Yudin and Ivanov}(2001)}]{Yudin_2001A}
\bibinfo{author}{\bibfnamefont{G.~L.} \bibnamefont{Yudin}} \bibnamefont{and}
  \bibinfo{author}{\bibfnamefont{M.~Y.} \bibnamefont{Ivanov}},
  \bibinfo{journal}{Phys. Rev. A} \textbf{\bibinfo{volume}{63}},
  \bibinfo{pages}{033404} (\bibinfo{year}{2001}).

\bibitem[{\citenamefont{Panfili et~al.}(2002)\citenamefont{Panfili, Haan, and
  Eberly}}]{Panfili_2002}
\bibinfo{author}{\bibfnamefont{R.}~\bibnamefont{Panfili}},
  \bibinfo{author}{\bibfnamefont{S.~L.} \bibnamefont{Haan}}, \bibnamefont{and}
  \bibinfo{author}{\bibfnamefont{J.~H.} \bibnamefont{Eberly}},
  \bibinfo{journal}{Phys. Rev. Lett.} \textbf{\bibinfo{volume}{89}},
  \bibinfo{pages}{113001} (\bibinfo{year}{2002}).

\bibitem[{\citenamefont{{Figueira de Morisson Faria} and
  Becker}(2003)}]{deMorissonFaria2003}
\bibinfo{author}{\bibfnamefont{C.}~\bibnamefont{{Figueira de Morisson Faria}}}
  \bibnamefont{and} \bibinfo{author}{\bibfnamefont{W.}~\bibnamefont{Becker}},
  \bibinfo{journal}{Laser Physics} \textbf{\bibinfo{volume}{13}},
  \bibinfo{pages}{1196} (\bibinfo{year}{2003}).

\bibitem[{\citenamefont{{Figueira de Morisson Faria}
  et~al.}(2004)\citenamefont{{Figueira de Morisson Faria}, Liu, Sanpera, and
  Lewenstein}}]{deMorissonFaria2004}
\bibinfo{author}{\bibfnamefont{C.}~\bibnamefont{{Figueira de Morisson Faria}}},
  \bibinfo{author}{\bibfnamefont{X.}~\bibnamefont{Liu}},
  \bibinfo{author}{\bibfnamefont{A.}~\bibnamefont{Sanpera}}, \bibnamefont{and}
  \bibinfo{author}{\bibfnamefont{M.}~\bibnamefont{Lewenstein}},
  \bibinfo{journal}{Phys. Rev. A} \textbf{\bibinfo{volume}{70}},
  \bibinfo{pages}{043406} (\bibinfo{year}{2004}).

\bibitem[{\citenamefont{Ho and Eberly}(2005)}]{Phay2005}
\bibinfo{author}{\bibfnamefont{P.~J.} \bibnamefont{Ho}} \bibnamefont{and}
  \bibinfo{author}{\bibfnamefont{J.~H.} \bibnamefont{Eberly}},
  \bibinfo{journal}{Phys. Rev. Lett.} \textbf{\bibinfo{volume}{95}},
  \bibinfo{pages}{193002} (\bibinfo{year}{2005}).

\bibitem[{\citenamefont{Becker and Faisal}(2005)}]{Becker2005a}
\bibinfo{author}{\bibfnamefont{A.}~\bibnamefont{Becker}} \bibnamefont{and}
  \bibinfo{author}{\bibfnamefont{F.~H.~M.} \bibnamefont{Faisal}},
  \bibinfo{journal}{J. Phys. B} \textbf{\bibinfo{volume}{38}},
  \bibinfo{pages}{R1} (\bibinfo{year}{2005}).

\bibitem[{\citenamefont{{Figueira de Morisson Faria} and
  Lewenstein}(2005)}]{deMorissonFaria2005}
\bibinfo{author}{\bibfnamefont{C.}~\bibnamefont{{Figueira de Morisson Faria}}}
  \bibnamefont{and}
  \bibinfo{author}{\bibfnamefont{M.}~\bibnamefont{Lewenstein}},
  \bibinfo{journal}{J. Phys. B} \textbf{\bibinfo{volume}{38}},
  \bibinfo{pages}{3251} (\bibinfo{year}{2005}).

\bibitem[{\citenamefont{Ho et~al.}(2005)\citenamefont{Ho, Panfili, Haan, and
  Eberly}}]{Ho_2005}
\bibinfo{author}{\bibfnamefont{P.~J.} \bibnamefont{Ho}},
  \bibinfo{author}{\bibfnamefont{R.}~\bibnamefont{Panfili}},
  \bibinfo{author}{\bibfnamefont{S.~L.} \bibnamefont{Haan}}, \bibnamefont{and}
  \bibinfo{author}{\bibfnamefont{J.~H.} \bibnamefont{Eberly}},
  \bibinfo{journal}{Phys. Rev. Lett.} \textbf{\bibinfo{volume}{94}},
  \bibinfo{pages}{093002} (\bibinfo{year}{2005}).

\bibitem[{\citenamefont{Ho and Eberly}(2006)}]{Ho2006}
\bibinfo{author}{\bibfnamefont{P.~J.} \bibnamefont{Ho}} \bibnamefont{and}
  \bibinfo{author}{\bibfnamefont{J.~H.} \bibnamefont{Eberly}},
  \bibinfo{journal}{Phys. Rev. Lett.} \textbf{\bibinfo{volume}{97}},
  \bibinfo{pages}{083001} (\bibinfo{year}{2006}).

\bibitem[{\citenamefont{Haan et~al.}(2006)\citenamefont{Haan, Breen, Karim, and
  Eberly}}]{Haan2006}
\bibinfo{author}{\bibfnamefont{S.~L.} \bibnamefont{Haan}},
  \bibinfo{author}{\bibfnamefont{L.}~\bibnamefont{Breen}},
  \bibinfo{author}{\bibfnamefont{A.}~\bibnamefont{Karim}}, \bibnamefont{and}
  \bibinfo{author}{\bibfnamefont{J.~H.} \bibnamefont{Eberly}},
  \bibinfo{journal}{Phys. Rev. Lett.} \textbf{\bibinfo{volume}{97}},
  \bibinfo{pages}{103008} (\bibinfo{year}{2006}).

\bibitem[{\citenamefont{{Figueira de Morisson Faria}
  et~al.}(2006)\citenamefont{{Figueira de Morisson Faria}, Liu, and
  Becker}}]{deMorissonFaria2006}
\bibinfo{author}{\bibfnamefont{C.}~\bibnamefont{{Figueira de Morisson Faria}}},
  \bibinfo{author}{\bibfnamefont{X.}~\bibnamefont{Liu}}, \bibnamefont{and}
  \bibinfo{author}{\bibfnamefont{W.}~\bibnamefont{Becker}},
  \bibinfo{journal}{J. Mod. Opt.} \textbf{\bibinfo{volume}{53}},
  \bibinfo{pages}{193} (\bibinfo{year}{2006}).

\bibitem[{\citenamefont{Liu et~al.}(2006)\citenamefont{Liu, {Figueira de
  Morisson Faria}, Becker, and Corkum}}]{Liu_2006}
\bibinfo{author}{\bibfnamefont{X.}~\bibnamefont{Liu}},
  \bibinfo{author}{\bibfnamefont{C.}~\bibnamefont{{Figueira de Morisson
  Faria}}}, \bibinfo{author}{\bibfnamefont{W.}~\bibnamefont{Becker}},
  \bibnamefont{and} \bibinfo{author}{\bibfnamefont{P.~B.}
  \bibnamefont{Corkum}}, \bibinfo{journal}{J. Phys. B}
  \textbf{\bibinfo{volume}{39}}, \bibinfo{pages}{L305} (\bibinfo{year}{2006}).

\bibitem[{\citenamefont{Taylor et~al.}(2007)\citenamefont{Taylor, Parker,
  Dundas, and Meharg}}]{Taylor2007}
\bibinfo{author}{\bibfnamefont{K.~T.} \bibnamefont{Taylor}},
  \bibinfo{author}{\bibfnamefont{J.~S.} \bibnamefont{Parker}},
  \bibinfo{author}{\bibfnamefont{D.}~\bibnamefont{Dundas}}, \bibnamefont{and}
  \bibinfo{author}{\bibfnamefont{K.~J.} \bibnamefont{Meharg}},
  \bibinfo{journal}{J. Mod. Opt.} \textbf{\bibinfo{volume}{54}},
  \bibinfo{pages}{1959} (\bibinfo{year}{2007}).

\bibitem[{\citenamefont{Staudte et~al.}(2007)\citenamefont{Staudte, Ruiz,
  Sch\"{o}ffler, Sch\"{o}ssler, Zeidler, Weber, Meckel, Villeneuve, Corkum,
  Becker et~al.}}]{Staudte_2007}
\bibinfo{author}{\bibfnamefont{A.}~\bibnamefont{Staudte}},
  \bibinfo{author}{\bibfnamefont{C.}~\bibnamefont{Ruiz}},
  \bibinfo{author}{\bibfnamefont{M.}~\bibnamefont{Sch\"{o}ffler}},
  \bibinfo{author}{\bibfnamefont{S.}~\bibnamefont{Sch\"{o}ssler}},
  \bibinfo{author}{\bibfnamefont{D.}~\bibnamefont{Zeidler}},
  \bibinfo{author}{\bibfnamefont{T.}~\bibnamefont{Weber}},
  \bibinfo{author}{\bibfnamefont{M.}~\bibnamefont{Meckel}},
  \bibinfo{author}{\bibfnamefont{D.~M.} \bibnamefont{Villeneuve}},
  \bibinfo{author}{\bibfnamefont{P.~B.} \bibnamefont{Corkum}},
  \bibinfo{author}{\bibfnamefont{A.}~\bibnamefont{Becker}},
  \bibnamefont{et~al.}, \bibinfo{journal}{Phys. Rev. Lett.}
  \textbf{\bibinfo{volume}{99}}, \bibinfo{pages}{263002}
  (\bibinfo{year}{2007}).

\bibitem[{\citenamefont{Rudenko et~al.}(2007)\citenamefont{Rudenko, de~Jesus,
  Ergler, Zrost, Feuerstein, Schr\"{o}ter, Moshammer, and
  Ullrich}}]{Rudenko_2007}
\bibinfo{author}{\bibfnamefont{A.}~\bibnamefont{Rudenko}},
  \bibinfo{author}{\bibfnamefont{V.~L.~B.} \bibnamefont{de~Jesus}},
  \bibinfo{author}{\bibfnamefont{T.}~\bibnamefont{Ergler}},
  \bibinfo{author}{\bibfnamefont{K.}~\bibnamefont{Zrost}},
  \bibinfo{author}{\bibfnamefont{B.}~\bibnamefont{Feuerstein}},
  \bibinfo{author}{\bibfnamefont{C.~D.} \bibnamefont{Schr\"{o}ter}},
  \bibinfo{author}{\bibfnamefont{R.}~\bibnamefont{Moshammer}},
  \bibnamefont{and} \bibinfo{author}{\bibfnamefont{J.}~\bibnamefont{Ullrich}},
  \bibinfo{journal}{Phys. Rev. Lett.} \textbf{\bibinfo{volume}{99}},
  \bibinfo{pages}{263003} (\bibinfo{year}{2007}).

\bibitem[{\citenamefont{{Figueira de Morisson Faria}
  et~al.}(2008)\citenamefont{{Figueira de Morisson Faria}, Shaaran, Liu, and
  Yang}}]{deMorissonFaria2008b}
\bibinfo{author}{\bibfnamefont{C.}~\bibnamefont{{Figueira de Morisson Faria}}},
  \bibinfo{author}{\bibfnamefont{T.}~\bibnamefont{Shaaran}},
  \bibinfo{author}{\bibfnamefont{X.}~\bibnamefont{Liu}}, \bibnamefont{and}
  \bibinfo{author}{\bibfnamefont{W.}~\bibnamefont{Yang}},
  \bibinfo{journal}{Phys. Rev. A} \textbf{\bibinfo{volume}{78}},
  \bibinfo{pages}{043407} (\bibinfo{year}{2008}).

\bibitem[{\citenamefont{Zhou et~al.}(2010)\citenamefont{Zhou, Liao, and
  Lu}}]{Zhou2010}
\bibinfo{author}{\bibfnamefont{Y.}~\bibnamefont{Zhou}},
  \bibinfo{author}{\bibfnamefont{Q.}~\bibnamefont{Liao}}, \bibnamefont{and}
  \bibinfo{author}{\bibfnamefont{P.}~\bibnamefont{Lu}}, \bibinfo{journal}{Phys.
  Rev. A} \textbf{\bibinfo{volume}{82}}, \bibinfo{pages}{053402}
  (\bibinfo{year}{2010}).

\bibitem[{\citenamefont{Chen et~al.}(2010)\citenamefont{Chen, Liang, and
  Lin}}]{Chen2010}
\bibinfo{author}{\bibfnamefont{Z.}~\bibnamefont{Chen}},
  \bibinfo{author}{\bibfnamefont{Y.}~\bibnamefont{Liang}}, \bibnamefont{and}
  \bibinfo{author}{\bibfnamefont{C.~D.} \bibnamefont{Lin}},
  \bibinfo{journal}{Phys. Rev. A} \textbf{\bibinfo{volume}{82}},
  \bibinfo{pages}{063417} (\bibinfo{year}{2010}).

\bibitem[{\citenamefont{Weckenbrock et~al.}(2004)\citenamefont{Weckenbrock,
  Zeidler, Staudte, Weber, Sch\"{o}ffler, Meckel, Kammer, Smolarski, Jagutzki,
  Bhardwaj et~al.}}]{Weckenbrock_2004}
\bibinfo{author}{\bibfnamefont{M.}~\bibnamefont{Weckenbrock}},
  \bibinfo{author}{\bibfnamefont{D.}~\bibnamefont{Zeidler}},
  \bibinfo{author}{\bibfnamefont{A.}~\bibnamefont{Staudte}},
  \bibinfo{author}{\bibfnamefont{T.}~\bibnamefont{Weber}},
  \bibinfo{author}{\bibfnamefont{M.}~\bibnamefont{Sch\"{o}ffler}},
  \bibinfo{author}{\bibfnamefont{M.}~\bibnamefont{Meckel}},
  \bibinfo{author}{\bibfnamefont{S.}~\bibnamefont{Kammer}},
  \bibinfo{author}{\bibfnamefont{M.}~\bibnamefont{Smolarski}},
  \bibinfo{author}{\bibfnamefont{O.}~\bibnamefont{Jagutzki}},
  \bibinfo{author}{\bibfnamefont{V.~R.} \bibnamefont{Bhardwaj}},
  \bibnamefont{et~al.}, \bibinfo{journal}{Phys. Rev. Lett.}
  \textbf{\bibinfo{volume}{92}}, \bibinfo{pages}{213002}
  (\bibinfo{year}{2004}).

\bibitem[{\citenamefont{Rudenko et~al.}(2004)\citenamefont{Rudenko, Zrost,
  Feuerstein, de~Jesus, Schr\"oter, Moshammer, and Ullrich}}]{Rudenko_2004}
\bibinfo{author}{\bibfnamefont{A.}~\bibnamefont{Rudenko}},
  \bibinfo{author}{\bibfnamefont{K.}~\bibnamefont{Zrost}},
  \bibinfo{author}{\bibfnamefont{B.}~\bibnamefont{Feuerstein}},
  \bibinfo{author}{\bibfnamefont{V.~L.~B.} \bibnamefont{de~Jesus}},
  \bibinfo{author}{\bibfnamefont{C.~D.} \bibnamefont{Schr\"oter}},
  \bibinfo{author}{\bibfnamefont{R.}~\bibnamefont{Moshammer}},
  \bibnamefont{and} \bibinfo{author}{\bibfnamefont{J.}~\bibnamefont{Ullrich}},
  \bibinfo{journal}{Phys. Rev. Lett.} \textbf{\bibinfo{volume}{93}},
  \bibinfo{pages}{253001} (\bibinfo{year}{2004}).

\bibitem[{\citenamefont{Zeidler et~al.}(2005)\citenamefont{Zeidler, Staudte,
  Bardon, Villeneuve, D{\"o}rner, and Corkum}}]{Zeidler_2005}
\bibinfo{author}{\bibfnamefont{D.}~\bibnamefont{Zeidler}},
  \bibinfo{author}{\bibfnamefont{A.}~\bibnamefont{Staudte}},
  \bibinfo{author}{\bibfnamefont{A.~B.} \bibnamefont{Bardon}},
  \bibinfo{author}{\bibfnamefont{D.~M.} \bibnamefont{Villeneuve}},
  \bibinfo{author}{\bibfnamefont{R.}~\bibnamefont{D{\"o}rner}},
  \bibnamefont{and} \bibinfo{author}{\bibfnamefont{P.~B.}
  \bibnamefont{Corkum}}, \bibinfo{journal}{Phys. Rev. Lett.}
  \textbf{\bibinfo{volume}{95}}, \bibinfo{pages}{203003}
  (\bibinfo{year}{2005}).

\bibitem[{\citenamefont{Zrost et~al.}(2006)\citenamefont{Zrost, Rudenko,
  Ergler, Feuerstein, de~Jesus, Schr\"oter, Moshammer, and
  Ullrich}}]{Zrost_2006}
\bibinfo{author}{\bibfnamefont{K.}~\bibnamefont{Zrost}},
  \bibinfo{author}{\bibfnamefont{A.}~\bibnamefont{Rudenko}},
  \bibinfo{author}{\bibfnamefont{T.}~\bibnamefont{Ergler}},
  \bibinfo{author}{\bibfnamefont{B.}~\bibnamefont{Feuerstein}},
  \bibinfo{author}{\bibfnamefont{V.~L.~B.} \bibnamefont{de~Jesus}},
  \bibinfo{author}{\bibfnamefont{C.~D.} \bibnamefont{Schr\"oter}},
  \bibinfo{author}{\bibfnamefont{R.}~\bibnamefont{Moshammer}},
  \bibnamefont{and} \bibinfo{author}{\bibfnamefont{J.}~\bibnamefont{Ullrich}},
  \bibinfo{journal}{J. Phys. B} \textbf{\bibinfo{volume}{39}},
  \bibinfo{pages}{S371} (\bibinfo{year}{2006}).

\bibitem[{\citenamefont{Liu et~al.}(2008)\citenamefont{Liu, Tschuch, Rudenko,
  D\"{u}rr, Siegel, Morgner, Moshammer, and Ullrich}}]{Liu2008a}
\bibinfo{author}{\bibfnamefont{Y.}~\bibnamefont{Liu}},
  \bibinfo{author}{\bibfnamefont{S.}~\bibnamefont{Tschuch}},
  \bibinfo{author}{\bibfnamefont{A.}~\bibnamefont{Rudenko}},
  \bibinfo{author}{\bibfnamefont{M.}~\bibnamefont{D\"{u}rr}},
  \bibinfo{author}{\bibfnamefont{M.}~\bibnamefont{Siegel}},
  \bibinfo{author}{\bibfnamefont{U.}~\bibnamefont{Morgner}},
  \bibinfo{author}{\bibfnamefont{R.}~\bibnamefont{Moshammer}},
  \bibnamefont{and} \bibinfo{author}{\bibfnamefont{J.}~\bibnamefont{Ullrich}},
  \bibinfo{journal}{Phys. Rev. Lett.} \textbf{\bibinfo{volume}{101}},
  \bibinfo{pages}{53001} (\bibinfo{year}{2008}).

\bibitem[{\citenamefont{Liu et~al.}(2010)\citenamefont{Liu, Ye, Liu, Rudenko,
  Tschuch, D\"{u}rr, Siegel, Morgner, Gong, Moshammer et~al.}}]{Liu2010}
\bibinfo{author}{\bibfnamefont{Y.}~\bibnamefont{Liu}},
  \bibinfo{author}{\bibfnamefont{D.}~\bibnamefont{Ye}},
  \bibinfo{author}{\bibfnamefont{J.}~\bibnamefont{Liu}},
  \bibinfo{author}{\bibfnamefont{A.}~\bibnamefont{Rudenko}},
  \bibinfo{author}{\bibfnamefont{S.}~\bibnamefont{Tschuch}},
  \bibinfo{author}{\bibfnamefont{M.}~\bibnamefont{D\"{u}rr}},
  \bibinfo{author}{\bibfnamefont{M.}~\bibnamefont{Siegel}},
  \bibinfo{author}{\bibfnamefont{U.}~\bibnamefont{Morgner}},
  \bibinfo{author}{\bibfnamefont{Q.}~\bibnamefont{Gong}},
  \bibinfo{author}{\bibfnamefont{R.}~\bibnamefont{Moshammer}},
  \bibnamefont{et~al.}, \bibinfo{journal}{Phys. Rev. Lett.}
  \textbf{\bibinfo{volume}{104}}, \bibinfo{pages}{173002}
  (\bibinfo{year}{2010}).

\bibitem[{\citenamefont{Haan et~al.}(2008)\citenamefont{Haan, Smith, Shomsky,
  and Plantinga}}]{Haan2008a}
\bibinfo{author}{\bibfnamefont{S.~L.} \bibnamefont{Haan}},
  \bibinfo{author}{\bibfnamefont{Z.~S.} \bibnamefont{Smith}},
  \bibinfo{author}{\bibfnamefont{K.~N.} \bibnamefont{Shomsky}},
  \bibnamefont{and} \bibinfo{author}{\bibfnamefont{P.~W.}
  \bibnamefont{Plantinga}}, \bibinfo{journal}{J. Phys. B}
  \textbf{\bibinfo{volume}{41}}, \bibinfo{pages}{211002}
  (\bibinfo{year}{2008}).

\bibitem[{\citenamefont{Emmanouilidou and Staudte}(2009)}]{Emmanouilidou2009}
\bibinfo{author}{\bibfnamefont{A.}~\bibnamefont{Emmanouilidou}}
  \bibnamefont{and} \bibinfo{author}{\bibfnamefont{A.}~\bibnamefont{Staudte}},
  \bibinfo{journal}{Phys. Rev. A} \textbf{\bibinfo{volume}{80}},
  \bibinfo{pages}{053415} (\bibinfo{year}{2009}).

\bibitem[{\citenamefont{Bondar et~al.}(2009{\natexlab{a}})\citenamefont{Bondar,
  Liu, and Ivanov}}]{Bondar2009b}
\bibinfo{author}{\bibfnamefont{D.~I.} \bibnamefont{Bondar}},
  \bibinfo{author}{\bibfnamefont{W.-K.} \bibnamefont{Liu}}, \bibnamefont{and}
  \bibinfo{author}{\bibfnamefont{M.}~\bibnamefont{Ivanov}},
  \bibinfo{journal}{Phys. Rev. A} \textbf{\bibinfo{volume}{79}},
  \bibinfo{pages}{023417} (\bibinfo{year}{2009}{\natexlab{a}}).

\bibitem[{\citenamefont{Shaaran and {Figueira de Morisson
  Faria}}(2010)}]{Shaaran2010}
\bibinfo{author}{\bibfnamefont{T.}~\bibnamefont{Shaaran}} \bibnamefont{and}
  \bibinfo{author}{\bibfnamefont{C.}~\bibnamefont{{Figueira de Morisson
  Faria}}}, \bibinfo{journal}{J. Mod. Opt.} \textbf{\bibinfo{volume}{57}},
  \bibinfo{pages}{984} (\bibinfo{year}{2010}).

\bibitem[{\citenamefont{Shaaran et~al.}(2010)\citenamefont{Shaaran, Nygren, and
  {Figueira de Morisson Faria}}}]{Shaaran2010a}
\bibinfo{author}{\bibfnamefont{T.}~\bibnamefont{Shaaran}},
  \bibinfo{author}{\bibfnamefont{M.~T.} \bibnamefont{Nygren}},
  \bibnamefont{and} \bibinfo{author}{\bibfnamefont{C.}~\bibnamefont{{Figueira
  de Morisson Faria}}}, \bibinfo{journal}{Phys. Rev. A}
  \textbf{\bibinfo{volume}{81}}, \bibinfo{pages}{063413}
  (\bibinfo{year}{2010}).

\bibitem[{\citenamefont{Ye and Liu}(2010)}]{Ye2010}
\bibinfo{author}{\bibfnamefont{D.~F.} \bibnamefont{Ye}} \bibnamefont{and}
  \bibinfo{author}{\bibfnamefont{J.}~\bibnamefont{Liu}},
  \bibinfo{journal}{Phys. Rev. A} \textbf{\bibinfo{volume}{81}},
  \bibinfo{pages}{043402} (\bibinfo{year}{2010}).

\bibitem[{\citenamefont{Haan et~al.}(2010)\citenamefont{Haan, Smith, Shomsky,
  Plantinga, and Atallah}}]{Haan2010}
\bibinfo{author}{\bibfnamefont{S.~L.} \bibnamefont{Haan}},
  \bibinfo{author}{\bibfnamefont{Z.~S.} \bibnamefont{Smith}},
  \bibinfo{author}{\bibfnamefont{K.~N.} \bibnamefont{Shomsky}},
  \bibinfo{author}{\bibfnamefont{P.~W.} \bibnamefont{Plantinga}},
  \bibnamefont{and} \bibinfo{author}{\bibfnamefont{T.~L.}
  \bibnamefont{Atallah}}, \bibinfo{journal}{Phys. Rev. A}
  \textbf{\bibinfo{volume}{81}}, \bibinfo{pages}{023409}
  (\bibinfo{year}{2010}).

\bibitem[{\citenamefont{Parker et~al.}(2006)\citenamefont{Parker, Doherty,
  Taylor, Schultz, Blaga, and DiMauro}}]{Parker2006}
\bibinfo{author}{\bibfnamefont{J.~S.} \bibnamefont{Parker}},
  \bibinfo{author}{\bibfnamefont{B.~J.~S.} \bibnamefont{Doherty}},
  \bibinfo{author}{\bibfnamefont{K.~T.} \bibnamefont{Taylor}},
  \bibinfo{author}{\bibfnamefont{K.~D.} \bibnamefont{Schultz}},
  \bibinfo{author}{\bibfnamefont{C.~I.} \bibnamefont{Blaga}}, \bibnamefont{and}
  \bibinfo{author}{\bibfnamefont{L.~F.} \bibnamefont{DiMauro}},
  \bibinfo{journal}{Phys. Rev. Lett.} \textbf{\bibinfo{volume}{96}},
  \bibinfo{pages}{133001} (\bibinfo{year}{2006}).

\bibitem[{\citenamefont{Dykhne}(1962)}]{Dykhne_1962}
\bibinfo{author}{\bibfnamefont{A.~M.} \bibnamefont{Dykhne}},
  \bibinfo{journal}{Sov. Phys. JETP} \textbf{\bibinfo{volume}{14}},
  \bibinfo{pages}{941} (\bibinfo{year}{1962}).

\bibitem[{\citenamefont{Solov'ev}(1976)}]{Solovev1976}
\bibinfo{author}{\bibfnamefont{E.~A.} \bibnamefont{Solov'ev}},
  \bibinfo{journal}{Sov. Phys. JETP} \textbf{\bibinfo{volume}{43}},
  \bibinfo{pages}{453} (\bibinfo{year}{1976}).

\bibitem[{\citenamefont{Solov'ev}(1989)}]{Solovev1989}
\bibinfo{author}{\bibfnamefont{E.~A.} \bibnamefont{Solov'ev}},
  \bibinfo{journal}{Sov. Phys. Uspekhi} \textbf{\bibinfo{volume}{32}},
  \bibinfo{pages}{228} (\bibinfo{year}{1989}).

\bibitem[{\citenamefont{Savichev}(1991)}]{Savichev1991a}
\bibinfo{author}{\bibfnamefont{V.~I.} \bibnamefont{Savichev}},
  \bibinfo{journal}{Sov. Phys. JETP} \textbf{\bibinfo{volume}{73}},
  \bibinfo{pages}{803} (\bibinfo{year}{1991}).

\bibitem[{\citenamefont{Solov'ev}(2005)}]{Solovev2005}
\bibinfo{author}{\bibfnamefont{E.~A.} \bibnamefont{Solov'ev}},
  \bibinfo{journal}{J. Phys. B} \textbf{\bibinfo{volume}{38}},
  \bibinfo{pages}{R153} (\bibinfo{year}{2005}).

\bibitem[{\citenamefont{Tolstikhin}(2008)}]{Tolstikhin2008a}
\bibinfo{author}{\bibfnamefont{O.~I.} \bibnamefont{Tolstikhin}},
  \bibinfo{journal}{Phys Rev. A} \textbf{\bibinfo{volume}{77}},
  \bibinfo{pages}{032711} (\bibinfo{year}{2008}).

\bibitem[{\citenamefont{Bondar et~al.}(2009{\natexlab{b}})\citenamefont{Bondar,
  Liu, and Yudin}}]{Bondar2009}
\bibinfo{author}{\bibfnamefont{D.~I.} \bibnamefont{Bondar}},
  \bibinfo{author}{\bibfnamefont{W.-K.} \bibnamefont{Liu}}, \bibnamefont{and}
  \bibinfo{author}{\bibfnamefont{G.~L.} \bibnamefont{Yudin}},
  \bibinfo{journal}{Phys. Rev. A} \textbf{\bibinfo{volume}{79}},
  \bibinfo{pages}{065401} (\bibinfo{year}{2009}{\natexlab{b}}).

\bibitem[{\citenamefont{Tolstikhin et~al.}(2010)\citenamefont{Tolstikhin,
  Morishita, and Watanabe}}]{Tolstikhin2010A}
\bibinfo{author}{\bibfnamefont{O.~I.} \bibnamefont{Tolstikhin}},
  \bibinfo{author}{\bibfnamefont{T.}~\bibnamefont{Morishita}},
  \bibnamefont{and} \bibinfo{author}{\bibfnamefont{S.}~\bibnamefont{Watanabe}},
  \bibinfo{journal}{Phys. Rev. A} \textbf{\bibinfo{volume}{81}},
  \bibinfo{pages}{033415} (\bibinfo{year}{2010}).

\bibitem[{\citenamefont{Batishchev et~al.}(2010)\citenamefont{Batishchev,
  Tolstikhin, and Morishita}}]{Batishchev2010}
\bibinfo{author}{\bibfnamefont{P.~A.} \bibnamefont{Batishchev}},
  \bibinfo{author}{\bibfnamefont{O.~I.} \bibnamefont{Tolstikhin}},
  \bibnamefont{and}
  \bibinfo{author}{\bibfnamefont{T.}~\bibnamefont{Morishita}},
  \bibinfo{journal}{Phys. Rev. A} \textbf{\bibinfo{volume}{82}},
  \bibinfo{pages}{023416} (\bibinfo{year}{2010}).

\bibitem[{\citenamefont{Bondar}(2010)}]{Bondar2010}
\bibinfo{author}{\bibfnamefont{D.~I.} \bibnamefont{Bondar}}, Ph.D. thesis,
  \bibinfo{school}{University of Waterloo} (\bibinfo{year}{2010}),
  \bibinfo{note}{available online at \url{http://hdl.handle.net/10012/5677} as
  well as at \href{http://arxiv.org/abs/1012.5334}{arXiv:1012.5334}.}

\bibitem[{\citenamefont{Bondar}(2008)}]{Bondar2008}
\bibinfo{author}{\bibfnamefont{D.~I.} \bibnamefont{Bondar}},
  \bibinfo{journal}{Phys. Rev. A} \textbf{\bibinfo{volume}{78}},
  \bibinfo{pages}{015405} (\bibinfo{year}{2008}).

\bibitem[{\citenamefont{Smirnova et~al.}(2008)\citenamefont{Smirnova, Spanner,
  and Ivanov}}]{Smirnova_2008}
\bibinfo{author}{\bibfnamefont{O.}~\bibnamefont{Smirnova}},
  \bibinfo{author}{\bibfnamefont{M.}~\bibnamefont{Spanner}}, \bibnamefont{and}
  \bibinfo{author}{\bibfnamefont{M.}~\bibnamefont{Ivanov}},
  \bibinfo{journal}{Phys. Rev. A} \textbf{\bibinfo{volume}{77}},
  \bibinfo{eid}{033407} (\bibinfo{year}{2008}).

\bibitem[{\citenamefont{Becker and Faisal}(1994)}]{Becker1994}
\bibinfo{author}{\bibfnamefont{A.}~\bibnamefont{Becker}} \bibnamefont{and}
  \bibinfo{author}{\bibfnamefont{F.~H.~M.} \bibnamefont{Faisal}},
  \bibinfo{journal}{Phys. Rev. A} \textbf{\bibinfo{volume}{50}},
  \bibinfo{pages}{3256} (\bibinfo{year}{1994}).

\bibitem[{\citenamefont{Becker and Faisal}(1995)}]{Becker1995}
\bibinfo{author}{\bibfnamefont{A.}~\bibnamefont{Becker}} \bibnamefont{and}
  \bibinfo{author}{\bibfnamefont{F.~H.~M.} \bibnamefont{Faisal}},
  \bibinfo{journal}{Phys. Rev. A} \textbf{\bibinfo{volume}{51}},
  \bibinfo{pages}{3390} (\bibinfo{year}{1995}).

\bibitem[{\citenamefont{Golovinski}(1997)}]{Golovinski1997}
\bibinfo{author}{\bibfnamefont{P.~A.} \bibnamefont{Golovinski}},
  \bibinfo{journal}{Laser Physics} \textbf{\bibinfo{volume}{7}},
  \bibinfo{pages}{655} (\bibinfo{year}{1997}).

\bibitem[{\citenamefont{Chirikov}(1979)}]{Chirikov1979}
\bibinfo{author}{\bibfnamefont{B.~V.} \bibnamefont{Chirikov}},
  \bibinfo{journal}{Phys. Rep.} \textbf{\bibinfo{volume}{52}},
  \bibinfo{pages}{263} (\bibinfo{year}{1979}).

\bibitem[{\citenamefont{Krainov}(2010)}]{Krainov2010}
\bibinfo{author}{\bibfnamefont{V.~P.} \bibnamefont{Krainov}},
  \bibinfo{journal}{JETP} \textbf{\bibinfo{volume}{111}}, \bibinfo{pages}{171}
  (\bibinfo{year}{2010}).

\end{thebibliography}
\end{document}